\documentclass[twocolumn,showpacs,superscriptaddress,floatfix,preprintnumbers,amssymb,amsmath]{revtex4}

\usepackage{graphicx}% Include figure files
\usepackage{dcolumn}% Align table columns on decimal point
\usepackage{bm}% bold math
\usepackage[latin1]{inputenc}
\usepackage[mathscr]{eucal}
\usepackage{epsfig}
\usepackage{rotating}

\begin{document}

\title{Experimental investigation of hybrid single-electron turnstiles with high charging energy}

\author{A.~Kemppinen}
\affiliation{Centre for Metrology and Accreditation (MIKES), P.O. Box 9, 02151 Espoo, Finland}
\author{S.~Kafanov}
\affiliation{Low Temperature Laboratory, Helsinki University of
Technology, P.O. Box 3500, 02015 TKK, Finland}
\author{Yu.~A.~Pashkin}
\affiliation{NEC Nano Electronics Research Laboratories and RIKEN Advanced Science Institute,
34 Miyukigaoka, Tsukuba, Ibaraki 305-8501, Japan}
\author{J.~S.~Tsai}
\affiliation{NEC Nano Electronics Research Laboratories and RIKEN Advanced Science Institute,
34 Miyukigaoka, Tsukuba, Ibaraki 305-8501, Japan}
\author{D.~V.~Averin}
\affiliation{Department of Physics and Astronomy, Stony Brook University, SUNY, Stony Brook, NY 11794-3800, USA}
\author{J.~P.~Pekola}
\affiliation{Low Temperature Laboratory, Helsinki University of
Technology, P.O. Box 3500, 02015 TKK, Finland}

\begin{abstract}
We present an experimental study of hybrid turnstiles with high charging energies in
comparison to the superconducting gap. The device is modeled with the sequential tunneling
approximation. The backtunneling effect is shown to limit the amplitude of the gate drive and
thereby the maximum pumped current of the turnstile. We compare results obtained
with sine and square wave drive and show how a fast rise time can suppress errors due to
leakage current. Quantized current plateaus up to 160~pA are demonstrated.
\end{abstract}


\maketitle Development of a quantum current standard based on the discreteness of the electron
charge has been one of the major goals of metrology for more than twenty
years~\cite{Averin1986}. The desired standard would generate the current $I=nef$ by transfering
$n$ electrons ($e$) per cycle at frequency $f$. The most precise device of this kind
was based on seven tunnel junctions in series~\cite{Keller1996}. It reached the relative 
accuracy of $10^{-8}$, but the maximum current was only on a picoampere level. However,
closing the quantum metrological triangle~\cite{Likharev1985}, which is one of the prime
needs for the quantum current standard, requires about 100~pA or more. The quest for
standards with higher currents has involved several candidates, based, e.g., on superconducting
devices~\cite{supra}, surface acoustic waves~\cite{Shilton1996},
and semiconducting quantum dots~\cite{Kaestner2009}. However, none of them
has reached metrological accuracy. In this Letter, we study the benefits of increased
charging energy for
the turnstile based on a hybrid single-electron transistor (SET), see Fig.~1(a),
proposed and demonstrated in Ref.~\cite{Pekola2008}. It was first realized as a NISIN
(N=normal-metal, I=insulator and S=superconductor) structure, but the SINIS version
reported in Ref.~\cite{Kemppinen2008} promises higher accuracy~\cite{Averin2008}.

The operation of the turnstile is based on the superconducting gap $\Delta$, which
expands the stability domains of the charge states, see Fig.~1(b-c). A small bias voltage
$V_\mathrm{b}$ can be applied since the current is ideally zero when
$|V_\mathrm{b}|<2\Delta/e$. The bias yields a preferred direction of tunneling, when
the gate voltage $V_\mathrm{b}(t)$ is alternated between the charge states. The gate voltage
is convenient to express in normalized units as
$V_\mathrm{g}(t)C_\mathrm{g}/e\equiv n_\mathrm{g}(t)\equiv n_\mathrm{g0}+A_\mathrm{g}w(t)$, where
$C_\mathrm{g}$, $n_\mathrm{g0}$ and $A_\mathrm{g}$ are the gate capacitance, dc offset
and drive amplitude, correspondingly. The gate waveform, e.g., sine or square, is expressed
with $w(t)$. By extending the gate sweep over several charge states, current plateaus with
$n>1$ can be obtained. However, the first plateau is optimal for metrology.

\begin{figure}[h]
    \begin{center}
    \includegraphics[width=.5\textwidth]{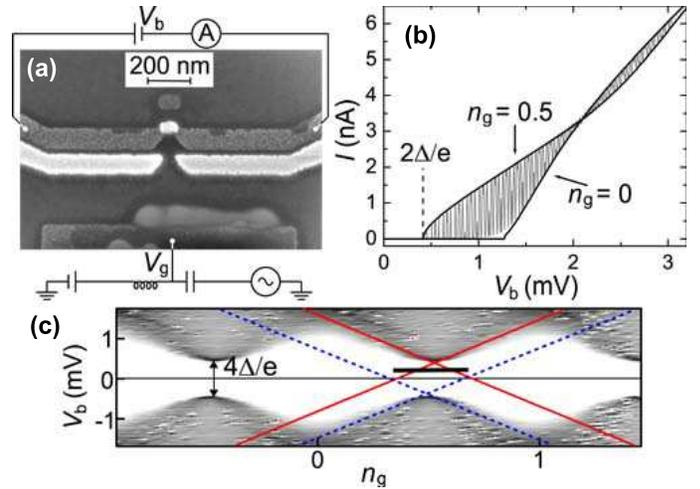}
    \end{center}
   \caption{\label{fig:iv} (a) Scanning electron micrograph of a typical turnstile, and
   a sketch of its measurement circuitry. The bright areas are AuPd (normal metal),
   and the darker parts are Al (superconductor).
   (b) Measured (grey) and simulated (black) dc IV curves of sample B. The dc gate
   voltage was swept during the measurement and hence the envelopes correspond to
   the simulated IV curves with $n_\mathrm{g}=0,0.5$.
   (c) Measured dynamic conductance $dI/dV_\mathrm{b}$ of sample A as a function
   of $V_\mathrm{b}$ and $n_\mathrm{g}$. The white areas are stability regions of
   charge states, where the conductance is negligible. They are limited by the
   thresholds for tunneling in the direction preferred by positive bias (solid lines) and
   negative bias (dashed lines). The pumping gate signal, when applied, alternates along the thick black
   line.}
\end{figure}

The errors of the turnstile were analyzed theoretically in Ref.~\cite{Averin2008} in which,
e.g., the tradeoff between the pumping speed and the error rate was considered.
Experimentally, the most critical error source appears to be the sub-gap leakage, i.e., residual
conductance in the superconducting gap~\cite{Pekola2008, Kemppinen2008, Lotkhov2009}. Often the
leakage shows up as a constant slope at
low bias. Then it can be described as the ratio $\eta_n = R_\mathrm{N}/R_n$ between the
normal-state resistance and the leakage resistance at plateau $n$. The index $n=0$ refers to
the dc zero-bias leakage. Higher-order tunneling events due to Andreev reflection or
Cooper-pair--electron cotunneling are known to be leakage processes that affect the operation of
the turnstile~\cite{Averin2008}. Possible other mechanisms include Cooper-pair
breaking~\cite{Dynes1984}, non-equilibrium quasiparticles and nonidealities of the tunnel barrier.
In Ref.~\cite{Averin2008} it was shown that the higher-order
processes are suppressed by the Coulomb blockade. Specifically, Andreev reflection vanishes in
the regime $n_\mathrm{g}=1/2\pm A_\mathrm{g,Ar}$, where $A_\mathrm{g,Ar}=1/2-\Delta/4E_\mathrm{c}$
is the threshold amplitude for Andreev reflection at the first plateau. The charging energy
$E_\mathrm{c}=e^2/2C_\Sigma$ depends on the total capacitance of the island, $C_\Sigma$. Note that
here and in the amplitude threshold discussion below we assume
the bias voltage $eV_\mathrm{b}=\Delta$, which minimizes the thermal errors~\cite{Pekola2008}.
Together with the threshold for the wanted forward tunneling process,
$A_\mathrm{g,ft}=\Delta/4E_\mathrm{c}$, $A_\mathrm{g,Ar}$ yields the requirement
$E_\mathrm{c}>\Delta$. In this case, the turnstile could be operated at error rates below
$10^{-8}$, which is the prime motivation for studying turnstiles with high charging energy. This
theoretical result was supported by the experiment of Ref.~\cite{Kemppinen2008}, where the leakage
$\eta_0$ was below $10^{-5}$. Yet, typically, the leakage of SINIS structures has been of
$\eta_0\sim 10^{-4}$. 

The first hybrid turnstiles were fabricated with standard two-layer electron beam lithography
process, and the charging energy was of the order of
$E_\mathrm{c}/k_\mathrm{B}\approx\Delta/k_\mathrm{B}\approx 2.5$~K~\cite{Pekola2008, Kemppinen2008}.
To increase $E_\mathrm{c}$, we used the trilayer germanium process~\cite{Pashkin2000}. The samples
were metallized in an electron-gun evaporator.

In this Letter, we present data from three samples, A, B and C, all with relatively high
charging energy, but with substantial differencies in $R_\mathrm{N}$.
All measurements were performed at temperature of about 0.1~K. The sample parameters are
summarized in Table~\ref{tab:param}. All samples show leakage that is too strong to be explained
by the higher-order processes, and that is strongest in the gate-open state, unlike expected
for Andreev reflection~\cite{Averin2008}. In addition, the leakage parameter $\eta_0$ does not
seem to depend on $R_\mathrm{N}$. Albeit counterproductive in general, the observable
leakage helps us here to demonstrate how the leakage at the plateau, $\eta_1$, can be suppressed
from $\eta_0$ by using square-wave drive.

\begin{table}[ht]
\caption{\label{tab:param} The parameters of the samples A, B and C.}
%\begin{small}
\begin{center}
\begin{tabular}{lccccc} \hline
 & $R_\mathrm{N}$ (k$\Omega$) & $E_\mathrm{c}/k_\mathrm{B}$ (K) & $\Delta$ ($\mu$eV) &
 $E_\mathrm{c}/\Delta$ & $\eta_0$ ($\times 10^{-5}$)\\ \hline
A & 1540 & 10 & 250 & 3.4 & 8\\
B & 315 & 7 & 210 & 2.9 & 4.7\\
C & 64 & 8.5 & 220 & 3.3 & 7\\ \hline
\end{tabular}
\end{center}
%\end{small}
\end{table}

The parameters of the samples were estimated by fitting their
dc IV curves to simulations based on the sequential tunneling model including the cooling
effect of the SINIS structure~\cite{Saira2007}. The measured and the simulated IV curves
of sample B are presented in Fig.~1(b).

The large resistance variation makes this set of samples ideal for studying the
speed of the turnstile. A high pumping frequency can cause two kinds of errors:
missed tunneling and backtunneling, i.e., tunneling in the direction
against the bias. When the gate signal is extended over the backtunneling thresholds,
the dashed lines in Fig.~1(c), tunneling in the forward direction is still preferred,
but backtunneling can be statistically significant. With sinusoidal drive and at low
frequencies, however, tunneling occurs typically before the backtunneling threshold is
exceeded. Note that backtunneling prevents precise pumping of
charge in the case of the normal metal SET~\cite{Averin2008}.

The backtunneling effect can be seen as backbending of the current plateaus in Fig.~2(a).
The effect is most pronounced at low $V_\mathrm{b}$ and high $A_\mathrm{g}$. The
experimental backtunneling slopes are reproduced precisely by our
simulations based on the sequential tunneling model with the sample parameters estimated
from independent dc measurements. Such an excellent agreement between the
experimental data and simulations was also observed for the other two samples.

\begin{figure}[h]
    \begin{center}
    \includegraphics[width=.5\textwidth]{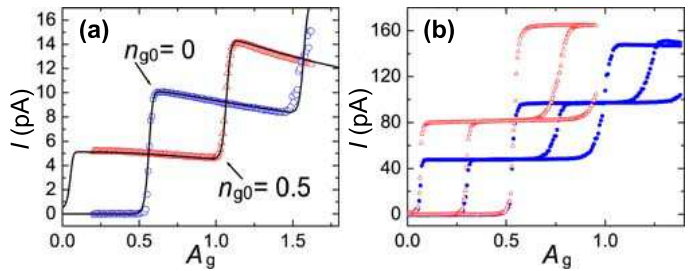}
    \end{center}
   \caption{\label{fig:bt} (a) The current of sample A vs.~gate amplitude with 32~MHz
   sine wave and at $eV_\mathrm{b}=0.8\Delta$. The lines show the corresponding simulated
   currents.
   (b) The measured plateaus of sample C with 300~MHz (solid circles) and
   500~MHz (open triangles) sinusoidal drives and at $n_\mathrm{g0}=0,0.25$ and 0.5.
   The bias voltage was $eV_\mathrm{b}=1.36\Delta$.}
\end{figure}

In Fig.~2(b), we demonstrate with the low-ohmic sample C a plateau at 160~pA,
which is an order of magnitude higher than in Refs.~\cite{Pekola2008,Kemppinen2008}. 
The experiments cannot be reproduced by our constant-temperature simulations, since there
is already significant heating at the highest
plateaus. More ideal behaviour is expected at the same current level at the first plateau.
However, we were limited by the maximum frequency of our rf generator, which is 500~MHz.

In Figs.~3(a-b) we show the effect of backtunneling on the pumped current in sample B
at the first plateau, $n_\mathrm{g0}=0.5$, and $eV_\mathrm{b}=\Delta$, with both sine and
square wave drive. In all cases, there is a flat region between the thresholds for forward
tunneling ($A_\mathrm{g,ft}$) and backtunneling ($A_\mathrm{g,bt}=3\Delta/4E_\mathrm{c}$),
shown by the dashed vertical lines. Above $A_\mathrm{g,bt}$, backtunneling
deteriorates the pumping accuracy. Since the rise time of the square wave was
constant, about 3~ns, the backtunneling rate did not depend on frequency.
At 120~MHz, sine and square waves are almost identical due to the finite rise time
of the square wave. 

High gate amplitude increases the rate of forward tunneling and thus decreases the effect
of missed tunneling. How close to $A_\mathrm{g,bt}$ the gate amplitude can be set,
depends on temperature. In general, the maximum amplitude is inversely proportional to
$E_\mathrm{c}$. In the experiments below, we used the amplitude half way between
the thresholds, $A_\mathrm{g,mean}=\Delta/2E_\mathrm{c}$, which is, e.g., about 0.17
for sample B. The pumping regime is thus drastically different from that of the first
turnstiles with $E_\mathrm{c}\approx \Delta$ and $A_\mathrm{g,mean}\approx 0.5$.

\begin{figure}[h]
    \begin{center}
    \includegraphics[width=.5\textwidth]{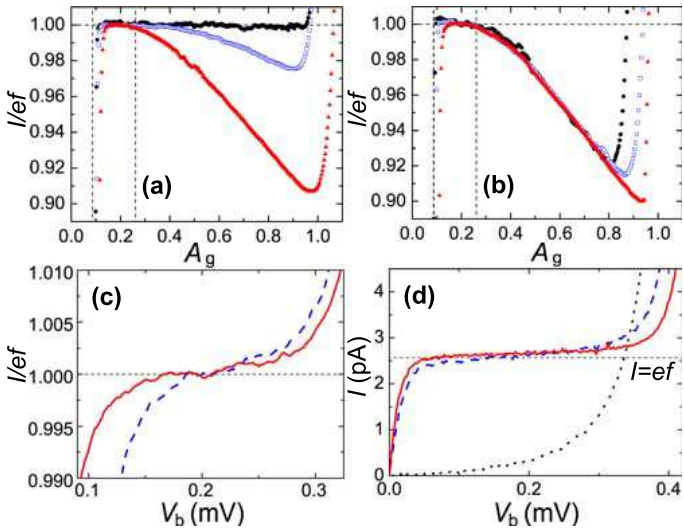}
    \end{center}
   \caption{\label{fig:squ} The first plateau of sample B at sine wave (a) and square wave
   (b). The frequencies are 120~MHz (solid triangles), 64~MHz (open squares), and 32~MHz
   (solid circles). $A_\mathrm{g,ft}$ and $A_\mathrm{g,bt}$ are marked
   with the dashed vertical lines. (c) The first plateau of
   sample B as a function of bias voltage at 64~MHz at sine wave (dashed line) and square
   wave (solid line). (d) The plateaus of sample C at 16~MHz sine (dashed line) and
   square (solid line) cases. For comparison, we show also the sub-gap dc IV curve with gate
   open (dotted line).}
\end{figure}

At zero temperature, the rate of single-electron tunneling process with electrostatic energy
change $E_p$ is $\Gamma_p=\sqrt{E_p^2-\Delta^2}/R_{\mathrm{N},i}e^2$, where $R_{\mathrm{N},i}$
is the resistance of the junction $i$. The
tunneling rates depend on $\Delta$, and since $E_p$ is roughly proportional to
$1/C_\Sigma$, they depend also on the $R_{\mathrm{N},i}C_\Sigma$ time constant.
However, if the gate amplitude is limited by the backtunneling effect, the forward
tunneling rate at the extreme gate value can be expressed as
$\Gamma_p=\gamma_\mathrm{a}\Delta/R_{\mathrm{N},i}e^2$. The tunneling rate is thus
independent of the capacitance. The prefactor $\gamma_\mathrm{a}$ depends on the exact
form of the amplitude limit, e.g., $\gamma_\mathrm{a}=\sqrt{5}/2$ for $A_\mathrm{g,mean}$
and $\gamma_\mathrm{a}=\sqrt{3}$ for $A_\mathrm{g,bt}$. In Ref.~\cite{Averin2008}, the
estimate for the maximum amplitude of the turnstile was based on $A_\mathrm{g,Ar}$,
which increases as a function of $E_\mathrm{c}$. Above
$E_\mathrm{c}=2\Delta$, however, backtunneling sets a more restrictive limit. Hence
the maximum current saturates to somewhat above 10~pA estimated for $E_\mathrm{c}=3\Delta$
with the superconducting gap of aluminum and error rate of $10^{-8}$.

There are two obvious strategies to increase the maximum current. Parallelization of
turnstiles is an attractive option because of the simplicity of the operation of a single
device. Another possibility is to find an alternative superconductor and to fabricate a
device with both high $\Delta$ and high $E_\mathrm{c}$. 

An additional benefit of high charging energy is that it suppresses the
leakage when the gate is not open. In Figs.~3(c-d) we demonstrate, as suggested in
Ref.~\cite{Averin2008}, how a square wave signal can be used to pass the gate-open state
quickly and hence to decrease the slope of the plateau as a function of bias voltage.
In Fig.~3(c), the plateau of sample B is clearly improved by using the square wave.

The effect of the square wave is more pronounced in the case of a fast sample and a low
frequency. Then the influence of leakage is high in relative terms, and the rise times
of the sine and square waves of our generator are significantly different. This can be
seen in Fig.~3(d) with sample C and frequency 16~MHz. The slopes with sine and square waves
are $\eta_1=9.8\times 10^{-5}$ and $\eta_1=3.8\times 10^{-5}$, correspondingly. Also,
both the sine and the square wave plateaus extend to higher bias voltage than the dc plateau.

The results of Fig.~3 are traceable to the national low-current standard of
MIKES~\cite{Iisakka2008}, which was used to calibrate the Keithley 6430 current meter.
It was then used to calibrate the DL Instruments 1211 current amplifier. 
The turnstile could not be measured with the Keithley current meter directly, because
of its relatively high input noise. The gain of the DL Instruments 1211 amplifier was
stable within few parts in $10^{-4}$ for several months. For each measurement, we
determined the input bias of the amplifier by comparing different polarities. The
uncertainty of our measurements was about $10^{-3}$. In the case of sample B at 64~MHz,
also the turnstile reached the accuracy of about $10^{-3}$.

To conclude, we have studied the operation of SINIS turnstiles with high charging energy.
The backtunneling effect was shown to limit the drive amplitude and thus the maximum current
of the turnstile at the highest charging energies. Under certain circumstances, a square
wave drive decreases the influence of sub-gap leakage. We note, however, that a square wave
tends to heat the island, whereas a sine signal can also cool it~\cite{Pekola2007}. Hence,
the optimum signal may be of some intermediate form.

We thank I. Iisakka for technical support and S.~Giblin, A.~Manninen, M.~Meschke and
M.~M\"{o}tt\"{o}nen for useful comments. The work was partially supported by Technology
Industries of Finland Centennial Foundation, the Academy of Finland and Japan Science
and Technology Agency through the CREST Project.
The research conducted within the EURAMET joint research project REUNIAM (A.K.)
and the EU project SCOPE (S.K. and J.P.) has received funding from the European
Community's Seventh Framework Programme under Grant Agreements No.~217257 and No.~218783.

\end{document}